\documentclass{sig-alternate}
\usepackage{mathptmx} 

\usepackage{paralist}
\usepackage{fancyhdr}
\usepackage[normalem]{ulem}
\usepackage[hyphens]{url}
\usepackage[sort,nocompress]{cite}
\usepackage[final]{microtype}
\usepackage[bookmarks=true,breaklinks=true,letterpaper=true,colorlinks,linkcolor=black,citecolor=blue,urlcolor=black]{hyperref}
\usepackage{multirow}
\usepackage{subfigure}
\usepackage{algorithm}
\usepackage{algorithmic}
\usepackage{enumitem}
\usepackage{comment}
\usepackage{wrapfig}
\usepackage{makecell} 

\fancypagestyle{firstpage}{
  \fancyhf{}
  
  \fancyhead[C]{\vspace{15pt}\normalsize{}} 
  \fancyfoot[C]{\thepage}
}

\title{GreenScale: Carbon-Aware Systems for Edge Computing 
\vspace{-1.2cm}
}

\author{
\normalsize{Young Geun Kim${^{\mathsection}}$}\thanks{{\scriptsize Correspondence to Young Geun Kim \(younggeun\_kim@korea.ac.kr\)}}  \;
\normalsize{Udit Gupta${^{\delta}}$} \;
\normalsize{Andrew McCrabb${^{\ddagger}}$} \;  \normalsize{Yonglak Son${^{\mathsection}}$} \; \normalsize{Valeria Bertacco${^{\ddagger}}$} \;
\normalsize{David Brooks${^{\dagger}}$${^{\delta}}$} \;\normalsize{Carole-Jean Wu${^{\delta}}$} \; \\ \\
\normalsize{${^{\mathsection}}$Korea University} \;\; \normalsize{${^{\dagger}}$Harvard University} \;\; \normalsize{${^{\ddagger}}$Univeristy of Michigan} \;\; \normalsize{${^{\delta}}$Meta}\\ \\
}

\begin{document}
\maketitle
\thispagestyle{firstpage}
\pagestyle{plain}


\begin{abstract}
To improve the environmental implications of the growing demand of computing, future applications need to improve the carbon-efficiency of computing infrastructures. State-of-the-art approaches, however, do not consider the intermittent nature of renewable energy. The \textit{time} and \textit{location}-based carbon intensity of energy fueling computing has been ignored when determining \textit{how} computation is carried out. This poses a new challenge --- deciding \textit{when} and \textit{where} to run applications across consumer devices at the edge and servers in the cloud. Such scheduling decisions become more complicated with the stochastic runtime variance and the amortization of the rising embodied emissions. This work proposes \textit{GreenScale}, a framework to understand the design and optimization space of carbon-aware scheduling for \textit{green applications} across the edge-cloud infrastructure. Based on the quantified carbon output of the infrastructure components, we demonstrate that optimizing for carbon, compared to performance and energy efficiency, yields unique scheduling solutions. Our evaluation with three representative categories of applications (i.e., AI, Game, and AR/VR) demonstrate that the carbon emissions of the applications can be reduced by up to 29.1\% with the \textit{GreenScale} --- with the scale of edge-cloud application users (in the order of millions), the yearly reduced carbon emission is 232.7$tCO_{2}$, on average, which is on par the average yearly carbon emissions of 55 vehicles. The analysis in this work further provides a detailed road map for edge-cloud application developers to build \textit{green applications}.
\end{abstract}

\section{Introduction}
\vspace{0.5cm}

With a dramatic advancement of information and communication technology (ICT), a variety of novel applications, such as artificial intelligence (AI), extended reality (XR), and cryptocurrencies, have been introduced~\cite{Amazon,Apple,Google_1,Google_2,Microsoft}. Despite the benefits of such applications, ICT has caused significant energy and environmental overheads worldwide. In 2022, the carbon emission from ICT accounts for 3\% of worldwide carbon emissions~\cite{AAndrea2015,UGupta2021}, which is on par with that from the aviation industry. It is even expected to account for up to 8\% of worldwide emissions in the next decade~\cite{UGupta2022}. There is a pressing need to design sustainable \textit{green applications} with minimal carbon.

\begin{figure}[t]
    \centering
    \includegraphics[width=\linewidth]{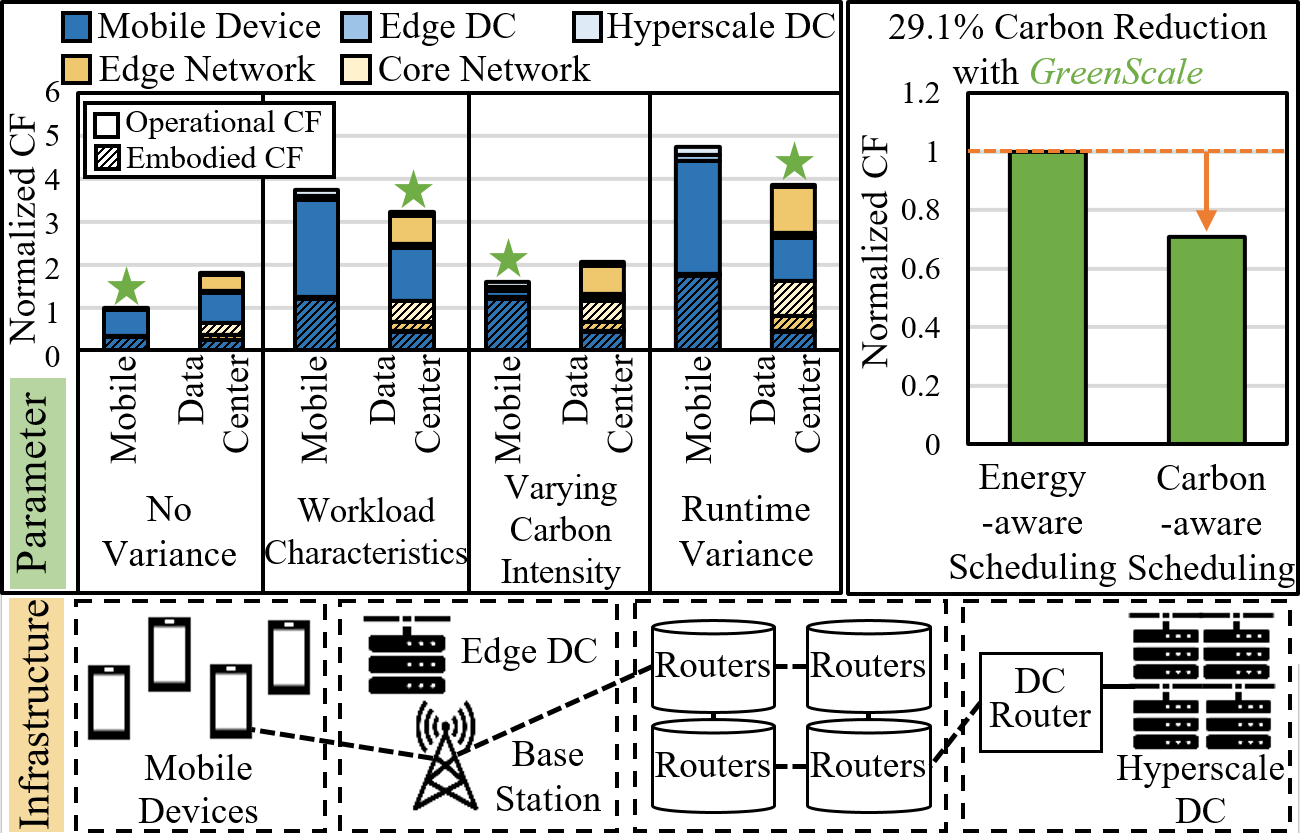}
    \vspace{-0.3cm}
    \caption{The overall carbon emission of the edge-cloud infrastructure significantly varies with workload characteristics, varying carbon intensity, and runtime variance. With judicious selections of execution targets for users using \textit{GreenScale}, the carbon emission can be significantly reduced by up to 29.1\%.}
    \label{fig:teaser}
    \vspace{-0.6cm}
\end{figure}

To design \textit{green applications}, developers will need to improve the carbon-efficiency of the infrastructure components. At-scale computing infrastructures have been significantly optimized by at-scale computing infrastructure design~\cite{Meta-DC,AWS-Efficiency,Google-DC}, operational efficiency, such as microprocessor energy efficiency and power usage effectiveness (PUE), optimization from industry~\cite{Google-TPU,Google-PUE,Meta-PUE,Meta-Zion} and academia~\cite{VZyuban2004,KKRangan2009,YLi2005}. Further operational energy efficiency improvement is increasingly more challenging~\cite{ECai2017,AEEshratifar2018,MHan2019,YKang2017,YKim2019,NDLane2016,SWang2020_1,SWang2020_2,GZhong2019,LWang2018,HYu2018,ASamantha2020}. In addition to efficiency optimization, renewable energy, such as solar and wind, is increasingly adopted to reduce computing's carbon footprint (CF)~\cite{STuli2022-2,ASouza2023,BAcun2023,Ryan2022}. Unfortunately, renewable energy generation is intermittent and is not always available at any single location all the time~\cite{BAcun2023,ASouza2023}. Due to the intermittent availability of renewable energy, the carbon intensity of computing across the computing spectrum of client devices at the edge and at-scale datacenter infrastructures can vary along with their locations and the time of use.

A \textit{new} challenge arises --- deciding \textit{where} to run applications \textit{when}, in order to minimize computing's carbon.
Modern computing infrastructures enable flexible computation execution via auto-scaling.
For example, many personalized AI and entertainment use cases are powered by a collaborative execution environment composed of smartphones and the cloud~\cite{CH,Fitbit,OMRON,WITHINGS}. In addition, virtual and augmented reality systems can consist of wearable electronics, smartphones as the staging device, and the cloud~\cite{Google_3, Google_4, Microsoft2, Oculus}. There are a variety of points where computations can occur. However, the decision process is challenging for any application, since the carbon intensity of each execution target across the edge-cloud infrastructure can significantly vary depending on \textit{where} and \textit{when} a computing execution target is charged or powered by \textit{what} energy generation sources.

To minimize an application's carbon, embodied emissions and runtime variance introduce additional challenges to the design of \textit{green applications}. Recent work characterizing the carbon footprint of modern hardware demonstrates that embodied carbon emissions (i.e., emissions owed to manufacturing processors, memory, and storage) are beginning to dominate computing's footprint~\cite{Dell,Google2019,Apple2019,UGupta2021,UGupta2022}, owing to high overheads of manufacturing integrated circuits and data center construction. Although it is not possible for an application to directly change the embodied carbon emissions, the embodied carbon emissions can be amortized by the users co-sharing the system during the application runtime.
Therefore, to maximize carbon efficiency, applications must take into account both operational \textit{and} embodied emissions when allocating users across the infrastructure components. Furthermore, mobile execution is stochastic by nature~\cite{BGaudette2016,BGaudette2019,YGKim2020}. Carbon-efficiency variability can stem from interference between and within applications~\cite{Wu:ISPASS11} and the stability of network~\cite{NDing2013}. Unfortunately, state-of-the-art approaches, such as~\cite{RMo2020,YZhang2018,Ryan2022,STuli2022-2,KKaur2020,YGKim2019-2,ECai2017,AEEshratifar2018,MHan2019,YKang2017,YKim2019,NDLane2016,SWang2020_1,SWang2020_2,GZhong2019,LWang2018,HYu2018,ASamantha2020}, determine the execution target primarily relying on the operational characteristics, such as performance and energy, without considering the aforementioned features (i.e., location- and time-dependent renewable energy availability, embodied emissions, and runtime variance), leaving a significant room for improvement (Fig.~\ref{fig:teaser}). This is due to the absence of a tool to quantify and analyze the carbon emissions of the infrastructure components by taking into account the aforementioned features, due to various challenges such as the heterogeneous interface across the system stacks and multiple organizations of infrastructure components. 

To enable insightful guidelines for developers to build \textit{green applications}, this paper proposes \textit{GreenScale} --- a carbon design space exploration and optimization framework.
\textit{GreenScale} quantifies the carbon emissions across the edge-cloud scheduling landscape by taking into account the application workload characteristics, location- and time-dependent renewable energy availability at power grids~\cite{electricityMap,Watttime}, amortization of embodied emissions~\cite{UGupta2022}, and runtime variance~\cite{BGaudette2016}. Based on the quantified carbon emissions, \textit{GreenScale} explores the design space of carbon-aware scheduling for three important edge-cloud application categories --- AI, Game, and AR/VR --- in different runtime environments.

The core contributions of this work include:
\begin{itemize}
    \item We build a framework, \textit{GreenScale}, to model and quantify the design space of carbon-aware \textit{green applications}. \textit{GreenScale} enables computation scaling with minimal environmental footprint considering the impact of application workload characteristics, location- and time-dependent renewable energy availability, amortization of embodied emissions, and runtime variance (Section~\ref{sec:design}). 
    \item By using \textit{GreenScale}, we explore the design space of carbon-aware scheduling for representative edge-cloud application categories. Based on the characterization results, we demonstrate \textit{carbon-aware scheduling} is distinct from conventional performance- and/or energy-aware scheduling, leaving a significant room for carbon improvement (Section~\ref{sec:result}).
    \item We additionally distill key insights for \textit{green application} development --- we provide guidelines for developers on making carbon-informed decisions for application design and system infrastructure parameters, such as, scheduling methods and resource provisioning (Section~\ref{sec:result4}) in order to minimize an application's carbon impact.
\end{itemize}
\section{Edge-Cloud Infrastructure}
\label{sec:background}
\begin{figure}[t!]
\includegraphics[width=\linewidth]{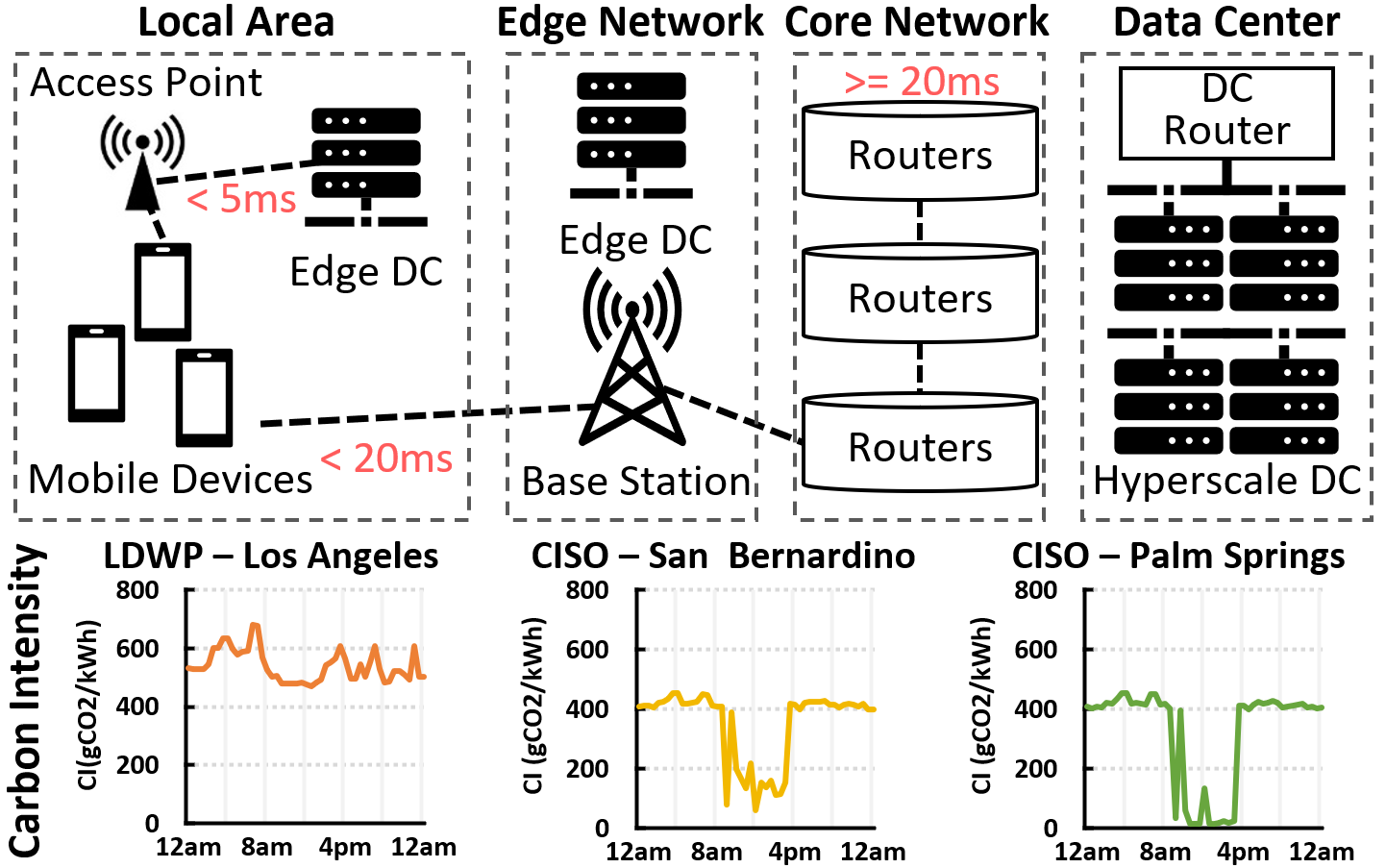}
\vspace{-0.4cm}
\caption{An example of edge-cloud infrastructure}
\label{fig:infrastructure}
\end{figure}

\vspace{0.5cm}
Applications aiming to run on the client devices at the edge can consider the distributed heterogeneous resources in the network hierarchy, from the edge of the network to the multi-hop remote cloud~\cite{RMo2020,YZhang2018,Ryan2022,STuli2022-2,KKaur2020}. Fig.~\ref{fig:infrastructure} shows an example of the edge-cloud infrastructure.

\textbf{Edge devices} are the computing resources located at the edge of the network (i.e., Local Area in Fig.~\ref{fig:infrastructure}). Examples of the edge devices include smartphones, smartwatches, laptops, appliances, and so forth. Traditionally, edge devices have been mainly used as user-end sensors, user interfaces, or both, in many mobile services. Recently, with the advancements in powerful mobile systems-on-a-chip (SoCs)~\cite{MHalpern2016,LNHuynh2017,SWang2020_2}, a varying amount of computations can be executed locally on the edge devices~\cite{CJWu2019,SWang2020_1,SWang2020_2,GZhong2019,LWang2018,HYu2018,ASamantha2020}. By doing so, the services can improve their response time and remove the data transmission overhead~\cite{ECai2017,AEEshratifar2018,MHan2019,YKang2017}. 

\begin{figure*}[t]
    \includegraphics[width=0.98\linewidth]{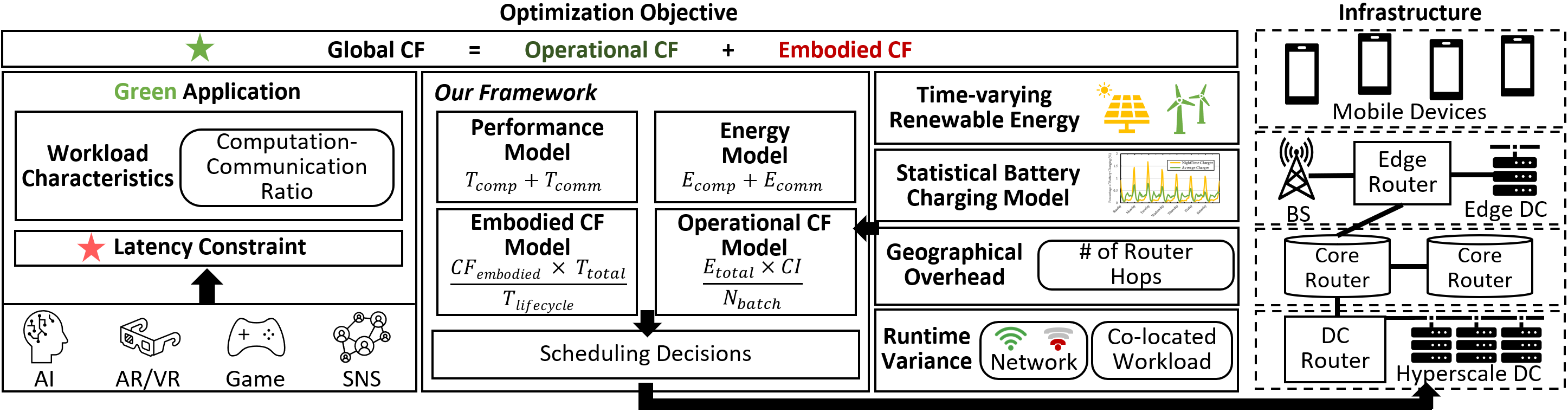} 
    \vspace{-0.2cm}
    \caption{Design space overview of carbon-aware green applications.}
    \label{fig:design_overview}
    \vspace{-0.3cm}
\end{figure*}

\textbf{Edge network} connects the edge devices in the multi-user access network with another access network, or the high-speed core network~\cite{MYan2019,DXu2020,AGupta2022}. For the mobile edge-cloud environment, the multi-user access network can be either a Wi-Fi access point network (i.e., Access Point in Local Area of Fig.~\ref{fig:infrastructure}) or a wireless cellular network (i.e., Edge Network in Fig.~\ref{fig:infrastructure}). In a typical cellular network, different scales (e.g., macro-, small-, and femto-scale) of base stations (BSs) are established in each cell, acting as a wireless transceiver~\cite{DXu2020}. 

\textbf{Edge data center} refers to the small-scale data center (DC) co-located within the edge network~\cite{DRichins2020,DRichins2021,Linux}. As shown in Fig.~\ref{fig:infrastructure}, edge DC can be located in either 1) urban area (e.g., in-house small-scale servers operated by the developer or the application company~\cite{BCharyyev2020}, or edge-scale cloud servers rented from the cloud service providers~\cite{EdgeConneX,365DataCenters,VaporIO}) or 2) rural area alongside the base station~\cite{Ericsson}. Since user-end devices only need to pass through the edge network, edge DCs can provide relatively powerful computing resources with tolerable data transmission overhead (from 5 ms to 20 ms depending on the location of edge DC as shown in Fig.~\ref{fig:infrastructure}). However, due to space and scalability limitations~\cite{Linux,KKaur2020}, the servers in an edge DC usually have less computing capabilities and efficiency than those in a hyperscale DC~\cite{BCharyyev2020}.

\textbf{Core network} includes the large-capacity core routers and high-speed fiber optic cables which connect consumers and businesses to DCs~\cite{MYan2019-2,MIorio2021,Cisco1,Cisco2}. 
Packets from edge networks are provided to a core router, passed from router to router, and delivered to another edge network which contains its destination. 

\textbf{Data center} refers to the large-scale DC usually operated by industrial companies~\cite{Google_1,Amazon-AWS,MS-Azure}. The DCs benefit from economies of scale for large workloads and co-location (i.e., batching) by exploiting a large number of highly efficient co-processors, such as GPUs and tensor processing units (TPUs). However, as the user-end devices need to pass through the edge network as well as the multiple hops of routers in the core network, DCs usually incur higher data transmission overheads compared to the edge DCs~\cite{RMo2020,YZhang2018,Ryan2022,STuli2022-2,KKaur2020}. 
\section{GreenScale}
\label{sec:design}
\vspace{0.4cm}


In this section, we propose \textit{GreenScale}, a carbon design space exploration and optimization framework to enable \textit{green application} development.
Fig.~\ref{fig:design_overview} shows the overall design space of carbon-aware edge-cloud scheduling.
The optimization objective of the design space is to minimize carbon emissions of an application, where computation kernels can be flexibly scheduled onto processors across the wide edge-cloud computing spectrum, satisfying the latency constraint. To achieve the optimization objective, \textit{GreenScale} quantifies the carbon emissions of infrastructure components for the computation kernels based on the performance and energy measurements with different application workloads under runtime variance, hourly energy generation data of all the US power grids powering the infrastructure components, and embodied CF modeling tools. Based on the quantified carbon emissions, \textit{GreenScale} exhaustively explores the carbon-optimal scheduling decisions in different environments.

To quantify the carbon emissions of infrastructure components, \textit{GreenScale} uses four models: 1) execution time performance model, 2) energy consumption model, 3) operational carbon footprint (CF) model, and 4) embodied CF model. Here, the performance and energy models are obtained based on the measurements with different application workloads under runtime variance. The operational CF model is calculated based on estimated energy consumption and carbon intensity of each infrastructure component~\cite{UGupta2021,UGupta2022,EIO-LCA,SImaPro,Sphera,Apple2019,Apple-R,TSMC,JChang2012}. For the embodied CF model, two embodied CF modeling tools are employed: ACT~\cite{UGupta2022} and LCA~\cite{EIO-LCA,SImaPro,Sphera,Apple-R,TSMC}. The details of the CF models are explained in Section~\ref{sec:design1}.

\subsection{Carbon Emission Model}
\label{sec:design1}
\vspace{0.5cm}
As explored by many previous works~\cite{UGupta2021,UGupta2022,EIO-LCA,SImaPro,Sphera,Apple2019,Apple-R,TSMC,JChang2012}, the carbon emissions across hardware life cycles can be split into four main phases: hardware manufacturing, hardware transport, operational use, and end-of-life processing and recycling. Among emissions from the phases, \textit{GreenScale} considers the operational carbon emissions along with embodied carbon emissions (i,e., emissions from the rest of hardware life cycles). Table~\ref{table:carbon_model} summarizes the carbon emission models of the infrastructure components and Table~\ref{table:carbon_notation} describes the abbreviations and notations used in the models.

\begin{scriptsize}
\begin{table*}[t]
  \caption{Carbon emission models of the infrastructure}
  \vspace{0.1cm}
  \centering
  \begin{tabular}{|l|l|l|l|}
    \hline
    \textbf{Execution Target} & \makecell{\textbf{Components}} & \makecell{\textbf{Operational CF}} & \makecell{\textbf{Embodied CF}} \\
    \hline
    \multirow{6}{2.1cm}{\makecell{Mobile Device}} &
    \makecell{Mobile Device} & \makecell{$T_{comp\_M} \times P_{comp\_M} \times CI_{M}$} & \makecell{$ECF_{M} \times \frac{T_{comp\_M}}{LT_{M}}$} \\ \cline{2-4}
    & \makecell{Edge Network} & \makecell{-} & \makecell{-} \\ \cline{2-4}
    & \makecell{Edge DC} & \makecell{$\frac{T_{comp\_M} \times P_{idle\_E\_DC}}{N_{user\_E}} \times CI_{E}$} &  \makecell{$\frac{ECF_{E\_DC}}{N_{user\_E}} \times \frac{T_{comp\_M}}{LT_{E\_DC}}$} \\ \cline{2-4}
    & \makecell{Core Network} & \makecell{-} & \makecell{-} \\ \cline{2-4}
    & \makecell{Hyperscale DC} & \makecell{$\frac{T_{comp\_M} \times P_{idle\_H}}{N_{user\_DC}} \times CI_{H}$} & \makecell{$\frac{ECF_{H}}{N_{user\_DC}} \times \frac{T_{comp\_M}}{LT_{H}}$} \\ \hline
    \multirow{8}{2.1cm}{\makecell{Edge DC}} &
    \makecell{Mobile Device} & \makecell{$(T_{comm\_E} \times P_{comm\_M} $ \\ $ + $ $ T_{comp\_E\_DC} \times P_{idle\_M}) \times CI_{M}$} & \makecell{$ECF_{M} \times \frac{(T_{comm\_E} + T_{comp\_E\_DC})}{LT_{M}}$} \\ \cline{2-4}
    & \makecell{Edge Network} & \makecell{$\frac{T_{comm\_E} \times P_{comp\_BS}}{N_{user\_BS}} \times CI_{E}$} & \makecell{$\frac{ECF_{BS}}{N_{user\_BS}} \times \frac{T_{comm\_E}}{LT_{BS}}$} \\ \cline{2-4}
    & \makecell{Edge DC} & \makecell{$\frac{T_{comp\_E\_DC} \times P_{comp\_E\_DC}}{N_{user\_E}} \times CI_{E}$} & \makecell{$\frac{ECF_{E\_DC}}{N_{user\_B}} \times \frac{T_{comp\_E\_DC}}{LT_{E\_DC}}$} \\ \cline{2-4}
    & \makecell{Core Network} & \makecell{-} & \makecell{-} \\ \cline{2-4}
    & \makecell{Hyperscale DC} & \makecell{$\frac{(T_{comm\_E} + T_{comp\_E\_DC}) \times P_{idle\_H}}{N_{user\_DC}} \times CI_{H}$} & \makecell{ $\frac{ECF_{H}}{N_{user\_DC}} \times \frac{(T_{comm\_E} + T_{comp\_E\_DC})}{LT_{H}}$} \\ \hline
    \multirow{8}{2.1cm}{\makecell{Hyperscale DC}} &
    \makecell{Mobile Device} &\makecell{$(T_{comm\_E} \times P_{comm\_M} $ \\ $ + $ $ (T_{comm\_R} + T_{comp\_H}) \times P_{idle\_M}) \times CI_{M}$} & \makecell{$ECF_{M} \times \frac{(T_{comm\_E} + T_{comm\_R} + T_{comp\_H})}{LT_{M}}$} \\ \cline{2-4}
    & \makecell{Edge Network} & \makecell{$\frac{T_{comm\_E} \times P_{comp\_BS}}{N_{user\_BS}} \times CI_{E}$} & \makecell{$\frac{ECF_{BS}}{N_{user\_BS}} \times \frac{T_{comm\_E}}{LT_{BS}}$ } \\ \cline{2-4}
    & \makecell{Edge DC} & \makecell{$\frac{(T_{comm\_E} + T_{comm\_R} + T_{comp\_H}) \times P_{idle\_E\_DC}}{N_{user\_E}} \times CI_{E}$} & \makecell{ $\frac{ECF_{E\_DC}}{N_{user\_E}} \times \frac{(T_{comm\_E} + T_{comm\_R} + T_{comp\_H})}{LT_{E\_DC}}$} \\ \cline{2-4}
    & \makecell{Core Network} & \makecell{$\frac{T_{comm\_R} \times P_{comm\_R}}{N_{user\_R}} \times CI_{R}$} & \makecell{ $\frac{ECF_{R}}{N_{user\_R}} \times \frac{T_{comm\_core}}{LT_{R}}$} \\ \cline{2-4}
    & \makecell{Hyperscale DC} & \makecell{$\frac{T_{comp\_H} \times P_{comp\_H}}{N_{B}} \times CI_{H}$} & \makecell{ $\frac{ECF_{H}}{N_{B}} \times \frac{T_{comp\_H}}{LT_{H}}$} \\ \hline
  \end{tabular}
  \vspace{-0.4cm}
  \label{table:carbon_model}
\end{table*}
\end{scriptsize}

\textbf{Operational Carbon Emission}: The operational carbon emission of each execution target is calculated based on the operational energy consumption of the components involved in the execution target and their respective carbon intensities ($CI$ in Table I)~\cite{BAcun2023,ASouza2023,UGupta2021,UGupta2022}. 

The operational energy consumption includes three components~\cite{YGKIM2015,LZhang2010,YGKim2017_1}: 1) computation energy, 2) communication energy, and 3) idle energy. The computation energy is the combination of the execution time ($T_{comp}$ in Table I) during which the required computations are performed on the actual execution target (either Mobile Device, Edge DC, or Hyperscale DC), and the power ($P_{comp}$ in Table I) consumed during the execution time. The communication energy is the combination of the time ($T_{comm}$ in Table I) during which the data is transmitted from the user-end device to the execution target and the power ($P_{comm}$ in Table I) consumed by each component during the data transmission time.
The idle energy consumption is the idle energy consumed by non-involved execution targets. The idle energy overhead is calculated based on the idle power ($P_{idle}$) consumed during the application runtime. It is divided by the number of users per components ($N_{user}$) assuming that the application developer (or company) along with users are responsible for the carbon emissions of the computing components that are used by the application.

Depending on which components are involved in the execution target, the operational energy consumption of the execution target varies. For example, when the mobile device is considered as the execution target, the computation energy consumption of mobile device as well as the idle energy overhead of the data centers are included in the operational energy consumption. On the other hand, when a data center is considered as the execution target, the communication energy of the client device and network components (i.e., Base Station or Core Routers), the computation energy of the data center, and the idle energy overhead of the client device are included in the operational energy.

The carbon intensity of each component depends on which energy source is used to power it. The carbon intensity of each energy source is summarized in Table~\ref{table:carbon_intensity}~\cite{BAcun2023,ASouza2023,UGupta2021,UGupta2022}. 
Typically, the components are powered by the closely located power grid~\cite{BAcun2023,ASouza2023,Watttime}. 
As hourly generation of energy sources in each grid highly depends on its location, the carbon intensity of each component is eventually determined depending on the location and time. To accurately model the carbon intensity of each component, the hourly energy generation data of all the US power grids~\cite{Watttime,electricityMap} are collected and used for \textit{GreenScale}.

\textbf{Embodied Carbon Emission}: The embodied carbon emission includes the emissions from the hardware life cycles except for the operational use. The embodied CF of a component is calculated based on its total embodied CF ($ECF$) and lifecycle time ($LT$), application runtime ($T_{comp}$ or $T_{comm}$), and the number of users co-sharing the component during the application runtime ($N_{user}$ or $N_{B}$), assuming that the embodied CF consumed during the application runtime (i.e., latency) can be discounted by the number of users co-sharing the component over the lifecycle time of the component. Here, $ECF$ of each component can be obtained by using the life cycle analysis tools (LCAs)~\cite{EIO-LCA,SImaPro,Sphera,Apple-R,TSMC} or the architectural carbon model tool (ACT)~\cite{UGupta2022}. Those models calculate the embodied CF based on a variety of parameters related with the hardware life cycles, such as SoC area, energy per area, gas per area, yield, raw materials, etc.

\subsection{Design Space Parameters}
\label{sec:design2}
\vspace{0.5cm}

\textbf{Workload characteristics: } The carbon optimal execution target depends on the workload characteristics. This is because the operational efficiency of the execution targets depends on the computation-communication ratio of the workloads. The computation energy and latency are determined by the amount of required computations of the workload (e.g., the number of floating point operations of AI workload), whereas the communication energy and latency are determined by the size of transmission data (e.g., input image or text of AI workload). Here, the energy is directly related with the operational carbon emission of the infrastructure components, and the latency is related with their embodied carbon emission.

The performance constraint is also dependent on the workload characteristics. For example, in case of AI workload, quality of service (QoS) expectations of users can be defined as a certain latency (e.g., 33.3 ms for 30 frames per second (FPS) video frame rate~\cite{BEgilmez2017,YZhu2015}, or 50 ms for interactive applications~\cite{YEndo1996,DLo2015}), below which users rarely perceive the difference. On the other hand, in case of game workload, more types of QoS metrics, such as latency, FPS, jitter, etc.~\cite{WCai2016,TLiu2020,MCarrascosa2022}, can be considered. Similar QoS metrics can also be used for AR/VR workloads~\cite{SZhao2021}. Since the QoS is a crucial metric for mobile optimization, it is also important to meet the workload-dependent performance constraints. 

\begin{table}[t]
  \caption{Abbreviations and Notations}
  \vspace{0.1cm}
  \centering
  \begin{tabular}{|l|l|}
    \hline
    \textbf{\makecell{Abbreviation \\ and Notation}} & \textbf{\makecell{Description}} \\ \hline
    $M$ & Mobile device \\ \hline
    $E$ & Edge \\ \hline
    $DC$ & Data center \\ \hline
    $H$ & Hyperscale DC \\ \hline
    $BS$ & Base station \\ \hline
    $R$ & Core Router \\ \hline
    $T_{comp}$ & Computation time \\ \hline
    $T_{comm}$ & Communication time \\ \hline
    $P_{comp}$ & Computation power consumption \\ \hline
    $P_{comm}$ & Communication power consumption \\ \hline
    $P_{idle}$ & Idle power consumption \\ \hline
    $CI$ & Carbon intensity \\ \hline
    $ECF$ & Total embodied carbon footprint \\ \hline
    $LT$ & Lifecycle time \\ \hline
    $N_{user\_DC}$ & Number of users per DC \\ \hline
    $N_{user\_edge}$ & Number of users at the edge system \\ \hline
    $N_{user\_BS}$ & Number of users per edge base station \\ \hline
    $N_{B}$ & Number of batched users in DC \\ \hline
  \end{tabular}
  \vspace{-0.6cm}
  \label{table:carbon_notation}
\end{table}

\textbf{Varying carbon intensity: } Another important factor that affects the carbon optimal execution target is the carbon intensity of energy sources used for the components. As we present in Table~\ref{table:carbon_intensity}, the carbon intensity of renewable energy (i.e., wind and solar) is much lower than that of the other energy sources, such as coal. Due to the carbon intensity gaps, the operational emissions of components significantly depend on the availability of energy sources, affecting the carbon optimal execution target. For example, in case of server-class computing systems whose operational emissions are larger than that of mobile systems, using renewable energy may significantly save the global carbon footprint, by reducing the carbon intensity of operational emissions.

\begin{figure}[t]
    \includegraphics[width=0.97\linewidth]{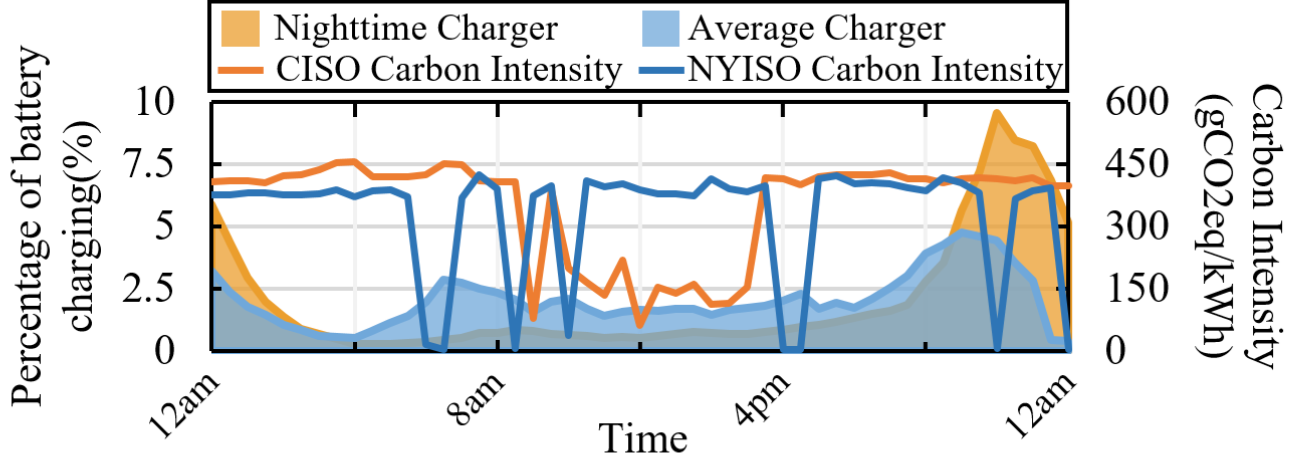} 
    \vspace{-0.4cm}
    \caption{Depending on when users charge their battery, the carbon intensity of the mobile device can vary; nighttime chargers mostly charge the battery when the carbon intensity is high (yellow area) whereas average chargers almost uniformly charge the battery (blue area).}
    \label{fig:user battery}
    \vspace{-0.5cm}
\end{figure}

However, the generation of each energy source in grids highly depends on their location and time~\cite{electricityMap,Watttime}. For example, in case of a grid in California (i.e., CISO in Fig.~\ref{fig:user battery}), the solar energy is only available during the daytime so that the carbon intensity of the grid is low only during the daytime. On the other hand, in case of a grid in New York state (i.e., NYISO in Fig.~\ref{fig:user battery}), the wind energy is available intermittently so that the carbon intensity of the grid fluctuates during a day. 

Due to the time-varying location-dependent energy source availability, the actual carbon intensity of the components can also vary. In case of the client devices at the edge, the carbon intensity can be determined when and how the users charge the battery of their devices~\cite{ASouza2023}. Fig.~\ref{fig:user battery} shows two different users' battery charging models along with the hourly carbon intensity of the grid in California (i.e., CISO) and that in New York State (NYISO): 1) nighttime charger who usually charges the battery during the nighttime and 2) average charger who charges the battery on demand throughout a day. As shown in Fig.~\ref{fig:user battery}, nighttime chargers may have higher carbon intensity for their mobile device since they mostly charges their battery when the carbon intensity is high. The carbon intensity may also depend on the location of the user.

In case of the data centers, the carbon intensity can be determined by the average carbon intensity of the powering grid, or the amount of renewable energy the companies purchase with sophisticated accounting frameworks that track renewable energy credits. Since the carbon optimal execution target can vary depending on the actual carbon intensity of the components, it is crucial to consider the time-varying location-dependent carbon intensity for the green application.

\begin{scriptsize}
\begin{table}[t]
  \caption{Operational carbon intensity of energy sources}
  \vspace{0.1cm}
  \centering
  \begin{tabular}{|l|l|l|l|}
    \hline
    \textbf{Type}& \textbf{$gCO_{2}eq/kWh$}\\
    \hline
    Wind & 11 \\
    \hline
    Solar & 41 \\
    \hline
    Water & 24 \\
    \hline
    Oil & 650 \\
    \hline
    Natural Gas & 490 \\
    \hline
    Coal & 820 \\
    \hline
    Nuclear & 12 \\
    \hline
    Other (Biofuels etc.) & 230 \\
    \hline
  \end{tabular}
  \label{table:carbon_intensity}
\end{table}
\end{scriptsize}

\textbf{Geographical trade-off: }Considering the time-varying renewable energy availability, it is also possible to consider other execution targets powered by another grid. For example, we can consider offloading to farther execution targets in spite of longer network latency (e.g., edge DC in rural area with a plenty of available renewable energy instead of mobile device or closer edge DC in Fig.~\ref{fig:infrastructure}), in order to have lower carbon intensity. Since the amount of trade-off between the network latency and carbon intensity may depend on the size of transmission data, it is important to carefully consider this geographical trade-off.

\textbf{Runtime variance: }The edge-cloud execution environment is stochastic by nature~\cite{BGaudette2016,BGaudette2019,YGKim2020}. In a realistic execution environment there can be several co-located workloads not only on the server-scale data centers but also on the mobile devices --- recent mobile devices support multi-tasking features, such as screen sharing of multiple applications. The resource interference from the co-located workloads can affect the computing efficiency of each components shifting the carbon optimal execution target. Network variability stemming from the varying wireless signal strength in edge network and congestion in core network can also significantly affect the communication efficiency. For example, the signal strength of wireless cellular network can vary considerably as edge device users move --- users undergo significant signal strength variations in daily life (43\% of data is transmitted under weak signal strength~\cite{NDing2013}). In addition, queuing delays in core networks are also dynamic depending on the traffic.
\section{Methodology}
\label{sec:methodology}
\vspace{0.2cm}

\subsection{Workloads}
\label{sec:methodology1}
\vspace{0.4cm}

To understand the carbon design space for green application with \textit{GreenScale}, we run three state-of-the-art workloads on our edge-cloud infrastructure: artificial intelligence (AI), game, and augmented/virtual reality (AR/VR). We explore the AI workloads as they are widely used in many of recent intelligent services and applications~\cite{MHan2019,VJReddi2019,wang2019exploiting,mattson2020mlperf}. In addition, we explore the game applications as they dominate 63.5\% of application market in 2022~\cite{Statista-app}. We also explore the AR/VR applications, as they have gained recent traction in both consumer and research communities~\cite{SZhao2021} thanks to the advances in efficient computing technologies, high-speed communication, and specialized hardware platforms.

\begin{scriptsize}
\begin{table}[t]
  \caption{NN inference workloads}
  \vspace{0.1cm}
  \centering
  \begin{tabular}{|c|c|c|c|c|}
    \hline
    \textbf{Category} & \textbf{NN} & \textbf{FLOPs} & \textbf{Params} & \textbf{\makecell{Input/ \\ Output \\ (KB)}} \\
    \hline
    & MobileNet & 0.31G & 3.5M & 150.5 \\ \cline{2-5} 
    & SqueezeNet & 0.82G & 1.2M & 150.5 \\ \cline{2-5}
    Vision& ResNet 50 & 4.09G & 25.6M & 150.5  \\ \cline{2-5}
    & MobileNet-SSD & 0.8G & 6.8M & 270 \\ \cline{2-5}
    & Inception & 5.71G & 23.8M & 268.2 \\ \hline
    Text & BERT & 25.3G & 17.5M & 1 \\ \hline
  \end{tabular}
 \vspace{-0.4cm}
  \label{table:DNN Workloads}
\end{table}
\end{scriptsize}

For AI workloads, we run the inference of state-of-the-art neural networks (NNs)~\cite{MHan2019,VJReddi2019,wang2019exploiting,mattson2020mlperf} which are widely used in mobile intelligent services~\cite{CH,Fitbit,OMRON,WITHINGS,Google_3, Google_4, Microsoft2, Oculus}. Table~\ref{table:DNN Workloads} summarizes the NNs. As shown in Table~\ref{table:DNN Workloads}, NNs have different computing characteristics depending on the number of floating point operations (FLOPs) and parameter sizes. In addition, they have different input/output sizes depending on the their categories. 
Different classes of NNs also have different performance requirements. For the vision NNs, we consider 30 FPS as the performance requirement --- users rarely perceive QoS difference as long as FPS exceeds 30~\cite{BEgilmez2017,YZhu2015}. For the text workload, we use 100 ms as the performance requirement~\cite{VJReddi2019}.

For game workloads, we run three different types of games summarized in Table~\ref{table:game Workloads} which are widely used by mobile users~\cite{WCai2016,TLiu2020,MCarrascosa2022}. As shown in Table~\ref{table:game Workloads}, different types of games have different performance requirements. For the case where the execution target is mobile devices, we directly run the Android applications on the mobile device while measuring latency, FPS, and power consumption of the device. On the other hand, for the case where the execution target is the DC, we run the cloud gaming Android application, NVIDIA Geforce Now~\cite{NVIDIA-G}, on the mobile device while measuring the latency, FPS, and power consumption of the device. For the power measurement of the cloud server, we also run the desktop version of the game on our DC infrastructure. 

For AR/VR workloads, we use four workloads from ILLIXR~\cite{MHuzaifa2021} summarized in Table~\ref{table:ar Workloads}: VR - 3D World~\cite{Monado}, VR - 3D Material~\cite{Godot1}, VR - 3D Cartoon~\cite{Godot2}, and AR~\cite{MHuzaifa2021}. AR/VR workloads have four sub tasks which include 1) Perception which reads inputs from sensors and understands the current surrounding environment, 2) Visual which combines the virtual information with physical world and renders the final frames, and 3) Audio which calculates and plays the audio. For all the workloads, the inputs are camera and motion tracking data and the outputs are the computed frames to be streamed on the client device.

\begin{scriptsize}
\begin{table}[t]
  \caption{Game workloads}
  \vspace{0.1cm}
  \centering
  \begin{tabular}{|c|c|c|c|c|}
    \hline
    \textbf{Category} & \textbf{Name} & \textbf{\makecell{Input/ \\ Output \\ (MB)}} & \textbf{\makecell{FPS \\ Req.}} & \textbf{\makecell{Latency \\ Req. (ms)}} \\
    \hline
    \makecell{1st-Person \\ Game} & Fortnite & 3.2 & 60 & 100  \\ \hline 
    \makecell{3rd-Person \\ Role Playing \\ Game} & \makecell{Genshin \\ Impact} & 3.0 & 60 & 500  \\ \hline
    \makecell{Omnipresent \\ Strategy} & \makecell{Team Fight \\ Tactics} &  1.9 & 60 & 1,000  \\ \hline
    \end{tabular}
    \vspace{-0.5cm}
  \label{table:game Workloads}
\end{table}
\end{scriptsize}

\begin{scriptsize}
\begin{table}[t]
  \caption{AR/VR workloads}
  \vspace{0.1cm}
  \centering
  \begin{tabular}{|c|c|c|c|c|}
    \hline
    \textbf{Category} & \textbf{Name} & \textbf{Tasks} & \textbf{\makecell{Input/ \\ Output \\ (KB)}} & \textbf{\makecell{Perf. \\ Req. \\ (ms)}} \\
    \hline
    \makecell{3D \\ World} & Sponza & \multirow{7}{1.8cm}{\makecell{1) Input \\ 2) Perception \\ 3) Visual/ \\ Audio}} & \multirow{7}{1cm}{\makecell{540.47}} & \multirow{7}{1cm}{\makecell{97.83}} \\ \cline{1-2}
    \makecell{3D \\ Material} & Materials & & &    \\ \cline{1-2}
    \makecell{3D \\ Cartoon} & Platformers &  & &  \\ \cline{1-2}
    AR & AR Demo & & & \\ \hline
    \end{tabular}
  \label{table:ar Workloads}
\end{table}
\end{scriptsize}

\subsection{Edge-Cloud Infrastructure}
\label{sec:methodology2}
\vspace{0.4cm}

This study for edge-cloud infrastructures is based on: mobile device, edge DC, DC, edge network, and core network. For AI and game workloads, we use an off-the-shelf Android smartphone, Pixel 3~\cite{Google2019}, as the mobile device, which is equipped with a Snapdragon 845 SoC~\cite{Qualcomm3}. On the other hand, for AR/VR workloads, we use an NVIDIA Jetson device which is equipped with NVIDIA Volta GPU, since the AR/VR workload does not run on Android devices. The specifications of the devices are summarized in Table~\ref{table:Devices}. We measure the power consumption of the devices by using an external power measurement device, Monsoon power monitor~\cite{Monsoon}.


For edge DC and hyperscale DC, we use the AWS instances of p3.2xlarge and p4d.24xlarge, respectively, which have similar specifications to the ones used in MLCommons~\cite{MLCommons,DRichins2020,DRichins2021}. For detailed specifications, see Table~\ref{table:specification}. 


For BS of edge network, we use the macro-scale BS whose TDP is 1000W~\cite{AGupta2022}. We calculate the single-user power consumption based on the number of users who can be served by the macro-scale BS~\cite{AGupta2022}. For the core routers, we use the typical power and bandwidth of off-the-shelf routers~\cite{Cisco1,Cisco2}.

\begin{table}[t]
\caption{Mobile device specification}
\vspace{0.1cm}
\centering
\begin{tabular}{|c|c|c|c|c|}
\hline
\textbf{Device}   & \textbf{CPU}           & \textbf{GPU}          & \textbf{Accel.} & \makecell{\textbf{RAM} \\ \textbf{(GB)}} \\ \hline
Pixel 3           & \makecell{Cortex A75\\2.4GHz}    & \makecell{Adreno 630\\0.7GHz}   & \makecell{Hexagon\\685}          & 4         \\ \hline
\makecell{Jetson\\AGX\\Xavier} & \makecell{NVIDIA\\Carmel\\2.4GHz} & \makecell{NVIDIA\\Volta\\1.4GHz} & NVDLA                & 64         \\ \hline
\end{tabular}
\label{table:Devices}
\end{table}

\subsection{GreenScale Parameters}
\vspace{0.4cm}

\textbf{Time-varying Location-Dependent Carbon Intensity: }As we explain in Section~\ref{sec:design1}, the carbon intensity of the components depends on \textit{when}, \textit{where}, and \textit{how} they are powered. To explore the realistic carbon intensity of the components, we consider reasonable scenarios along with the hourly renewable generation reports of the US grids~\cite{electricityMap,Watttime}.

For mobile devices, we consider three scenarios for user charging behaviors based on the previous user studies~\cite{DFerreira2011,EAOliver2011,SSaxena2017}: 1) nighttime charger who charges the battery only during the nighttime, 2) average charger who uniformly charges the battery throughout a day, and 3) intelligent charger who only charges the battery when the renewable energy is available at the close grid. By exploiting the statistical models in~\cite{DFerreira2011,EAOliver2011,SSaxena2017}, we calculate the average carbon intensity of the mobile device in different locations.

For edge data centers and base stations, we consider two scenarios depending on their locations~\cite{AGupta2022,DXu2020}: 1) urban area where the amount of renewable energy generation is relatively small (with short data transmission latency) and 2) rural area where there exist a plenty of renewable energy sources (with relatively longer data transmission latency). In case of the core routers, we use the average carbon intensity of the grids in any state, since the core routers are usually scattered across the state for delivering packets~\cite{joseph2009CiscoQoS,Cisco1,Cisco2}.

For data centers, we consider two scenarios depending on how they are powered~\cite{BAcun2023,ASouza2023}: 1) grid-mix where the data center is powered by the mixed energy sources generated by the closest grid and 2) carbon free\footnote{Note NetZero claimed by cloud providers (by annually offsetting the DC's energy with renewable energy credits) does not meas their actual energy is carbon free~\cite{BAcun2023} --- DCs continue to consume carbon-intensive energy when renewable energy supply is insufficient hourly.} where the data center is fully covered by the renewable energy.

\textbf{Runtime Variance: }We explore two types of runtime variance which users widely experience while using the applications~\cite{BGaudette2016,BGaudette2019,YGKim2020}: interference from co-located workload and network stability. For the interference from co-located workloads, we measure the computation latency on each component, while running other workloads. To model the unstable edge network, we measure the data transmission latency under the bad wireless signal strength~\cite{YGKim2017_1}, \cite{YGKim2020}. Similarly, we model the unstable core network based on the impact of variability (e.g., congestion) modeled in~\cite{MIorio2021,bauknecht2020investigation,OSPFiberOpticsGuide,ITUFiberCableInstallation}.

\begin{figure*}[t!]
\includegraphics[width=0.95\linewidth]{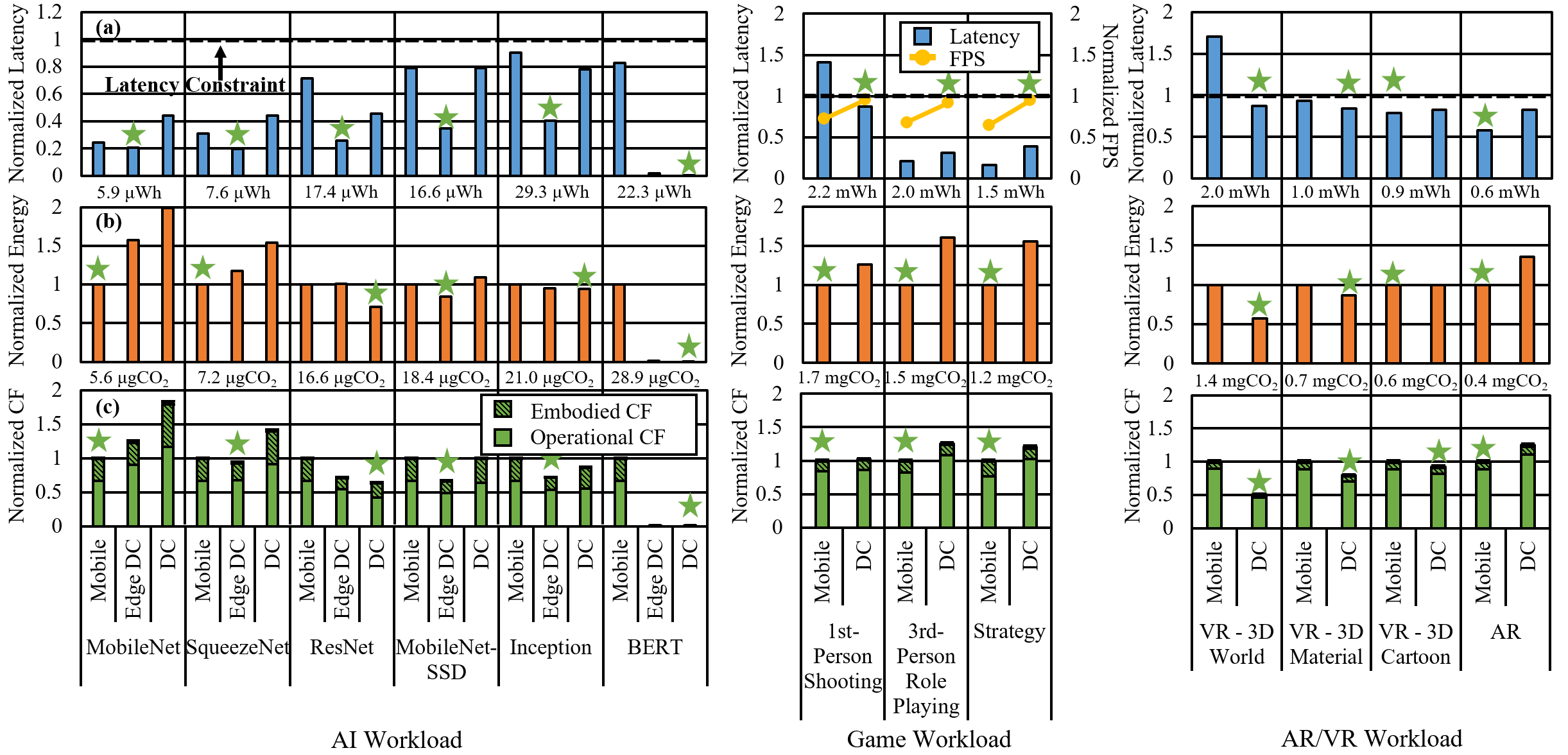}
\vspace{-0.3cm}
\caption{Carbon optimization design space across edge-cloud infrastructure: the performance-, energy-, and carbon-optimal execution targets vary depending on use cases. Note latency is normalized to the respective latency constraints of the workloads, whereas the energy and CF are normalized to those of Mobile for each workload --- numbers above (b) and (c) show the energy and CF of Mobile for each workload, respectively.}
\label{fig:result_overview}
\end{figure*}

\textbf{Embodied CF Models: }Although rising embodied carbon emissions have received much attention recently~\cite{UGupta2021,UGupta2022}, the deployment of embodied carbon footprint (CF) models is still in a nascent stage. To explore the impact of embodied CF models on the scheduling decisions of green applications, we use two embodied CF modeling tools: ACT and LCA tools. 

For the computing components, we first calculate the embodied carbon emissions of the computing components using ACT~\cite{UGupta2022}. We also obtain their embodied carbon emissions based on the LCA reports~\cite{EIO-LCA,SImaPro,Sphera,Apple-R,TSMC} provided by manufacturing companies. For example, in case of mobile devices, we use embodied CF presented in the sustainability report of Google Pixel 3~\cite{Google2019}. On the other hand, in case of server-class systems, we use embodied CF presented in the sustainability report of Dell R740~\cite{Dell}. Note, according to~\cite{UGupta2021}, those two modeling tools have 28\% gap in terms of the estimated embodied carbon emissions.

For the base station and routers, we do not use ACT, as the networking components, such as transceivers, are not modeled in ACT. We instead obtain their embodied carbon emissions based on the LCA reports provided by network infrastructure companies~\cite{Ericsson,Ericsson-1,Ericsson-2,Ericsson-3,Ericsson-4,Cisco1,Cisco2}.

\begin{table}[t]
\caption{Server specification}
\vspace{0.1cm}
\centering
\begin{tabular}{|c|c|c|c|}
\hline
\textbf{Type} & \textbf{CPU}          & \textbf{Accel.}   & \textbf{RAM} \\ \hline
p3.2xlarge    & \makecell{Intel Xeon\\E5-2686 v4} & \makecell{NVIDIA\\Tesla V100} & 64GB         \\ \hline
p4d.24xlarge  & \makecell{Intel Xeon\\P-8275CL}   & \makecell{NVIDIA\\A100 x8}    & 1152GB    \\ \hline
\end{tabular}
\label{table:specification}
\end{table}
\section{Evaluation Results and Analysis}
\label{sec:result}
\vspace{0.5cm}
In this section, we explore the design space of carbon-aware scheduling for representative edge-cloud applications by using \textit{GreenScale}. The design space includes 1) scheduling decisions (i.e., allocation of user requests across edge-cloud infrastructure) and 2) resource allocations (i.e., adjustments of the number of server instances to rent). For each application category, the size of design space is $\sim$200K, which is combination of workload characteristics, varying carbon intensity, runtime variance, and available scheduling decisions.

\subsection{Result Overview of Carbon Design Space}
\label{sec:result1}
\vspace{0.4cm}

Fig.~\ref{fig:result_overview} shows (a) the performance, (b) energy consumption, and (c) carbon footprint results for the carbon optimization design space for a variety of application use cases across the edge-cloud infrastructure. The x-axis shows the available execution targets for each workload whereas the y-axis of (a), (b), and (c) shows the normalized latency, energy, and carbon footprint of the execution targets, respectively. Here, the carbon intensity models of mobile device, edge DC, and hyperscale DC are Nighttime Charger, Urban Area, and Grid-Mix, respectively. ACT is used as the embodied CF modeling tool, and there is no runtime variance.

\vspace{0.1cm}
\noindent \textit{The carbon-optimal execution target is not always the same as the performance- or energy-optimal execution target. The performance-, energy-, and carbon-optimal execution target (green stars of Fig.~\ref{fig:result_overview}) varies based on the use cases}.
\vspace{0.1cm}

In case of the AI workloads, the latency-, energy-, and carbon-optimal execution targets depend on the computation-communication ratio of NNs. For example, although MobileNet, SqueezeNet, and ResNet have the same size of transmission data, the carbon-optimal execution target differs (i.e., Mobile, Edge DC, and DC, respectively), due to the different number of FLOPs. On the other hand, the carbon optimal-execution targets for MobileNet-SSD and Inception are Edge DC, as they have the larger data transmission overhead along with a higher number of FLOPs than MobileNet. In case of BERT, the latency-, energy-, and carbon-optimal execution targets are DC, due to the smallest data transmission overhead.

In case of the game workloads, DC (i.e., cloud gaming service) provides better quality of experience (QoE) compared to Mobile (i.e., Android application) by providing higher FPS satisfying latency constraint. However, due to the high data transmission overhead in terms of operational energy, Mobile always shows better carbon emission compared to DC --- DC needs to keep transmitting the captured frames to Mobile. 

In case of the AR/VR workloads, the latency-, energy- and carbon-optimal execution targets also depend on the computation-communication ratio of the workloads. In fact, all the AR/VR workloads have the same size of the transmission data as summarized in Table~\ref{table:ar Workloads}. However, VR - 3D World requires higher amount of computations compared to the other workloads. Due to the reason, it does not satisfy the latency constraint with Mobile, eventually having lower carbon footprint with DC. On the other hand, the rest of the workloads exhibit lower carbon footprint with Mobile. 

\vspace{0.1cm}
\noindent \textit{This work on GreenScale demonstrates the potential of an intelligent carbon-aware computation scheduling. An application's carbon cost can be reduced by 29.1 \% by scaling the scope of computations beyond local, client devices.}
\vspace{0.1cm}

\begin{figure}[t]
    \includegraphics[width=0.98\linewidth]{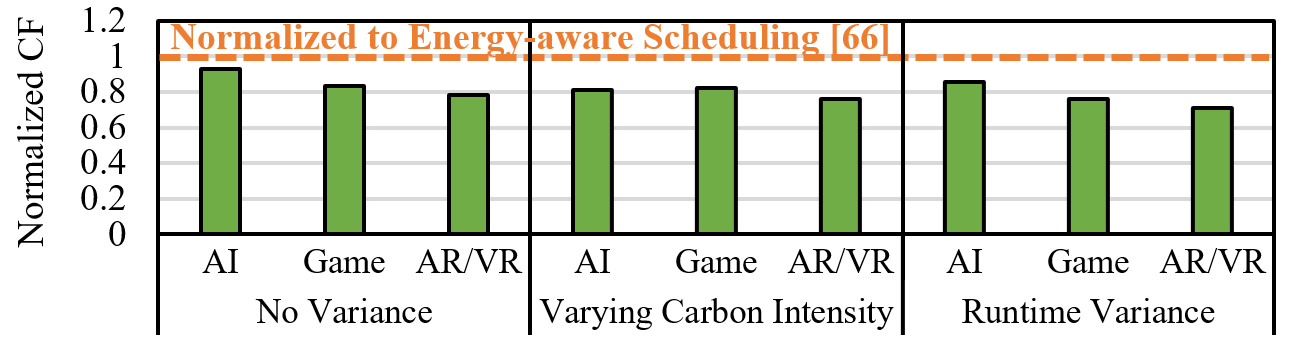} 
    \vspace{-0.25cm}
    \caption{Carbon optimization is distinct from performance and energy optimization leaving a significant room for carbon improvement. The x-axis shows design parameters and the y-axis shows the normalized CF of carbon-aware scheduling over energy-aware scheduling~\cite{YGKim2020}.}
    \label{fig:previous}
\end{figure}

The carbon optimization design space becomes even more complex in the presence of varying carbon intensity of the components and runtime variance. Due to the complex design space, the carbon optimization is distinct from the conventional performance and energy optimizations, leaving a significant room for carbon-efficiency improvement --- carbon-aware scheduler (i.e,. a scheduler that allocates user requests across components explicitly considering the carbon features) outweighs the state-of-the-art scheduler~\cite{YGKim2020} by up to 29.1\% (rightmost bar in Fig.~\ref{fig:previous}).

\begin{figure}[t]
    \includegraphics[width=0.98\linewidth]{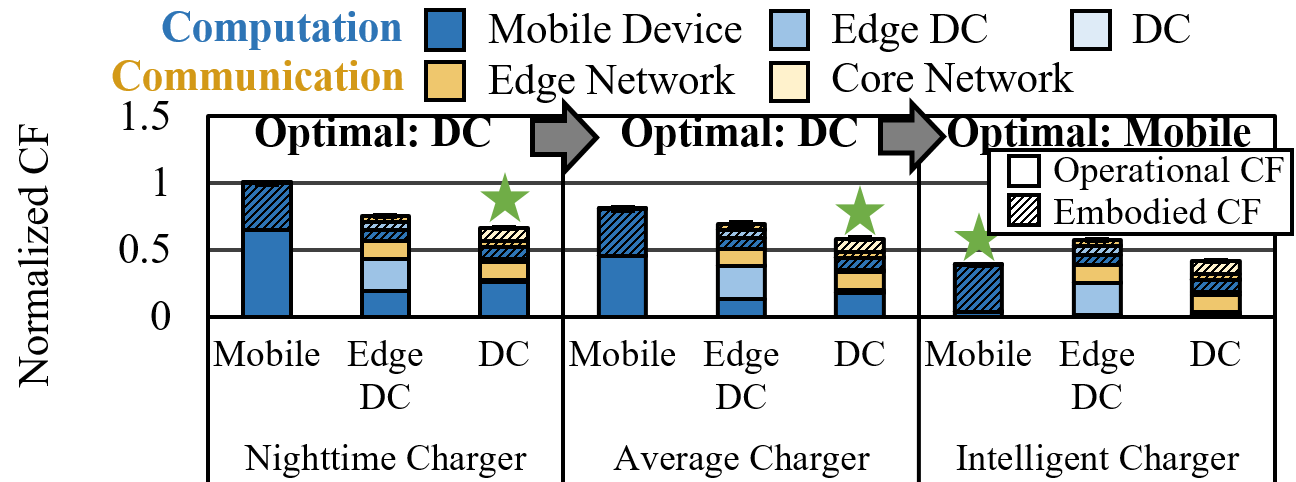} 
    \vspace{-0.25cm}
    \caption{CF of ResNet with different charging scenarios for Mobile. Note CF of execution targets is normalized to that of Mobile in Nighttime Charger. The carbon intensity of Mobile depends on when and how the mobile device user charges the battery. If the user intelligently charges the battery considering the renewable energy availability at the closest grid (i.e., Intelligent Charger), he/she can significantly save the carbon footprint (by 61.2\%).}
    \label{fig:user CI}
\end{figure}

\subsection{Impact of Varying Carbon Intensity}
\label{sec:result2}
\vspace{0.4cm}

\textbf{Carbon intensity of Mobile: }Fig.~\ref{fig:user CI} shows the CF of ResNet with different charging scenarios for Mobile. The x-axis shows the three execution targets with different battery charging scenarios (i.e., Nighttime Charger, Average Charger, and Intelligent Charger in each column) and the y-axis shows the cumulative CF of each execution target. Note, the carbon intensity models of edge DC and hyperscale DC are Urban Area and Grid-Mix, respectively. ACT is used as the embodied CF modeling tool, and there is no runtime variance.

\vspace{0.1cm}
\noindent \textit{Intelligent carbon-aware battery charging for client devices achieves up to 61.2\% carbon reduction. With the scale of client devices, from smartphones to wearables, in the order of billions~\cite{Statista}, the carbon impact is significant.}
\vspace{0.1cm}

In case of the Nighttime Charger, the carbon intensity of Mobile is usually high since the battery is charged during the nighttime where less amount of renewable energy is available. When the charging behavior of the user changes from nighttime charger to intelligent charger, the carbon intensity of the Mobile significantly decreases, saving the overall carbon footprint by up to 61.2\% (i.e., Mobile bars) --- note other workloads show similar result trends. In this case, the carbon optimal execution target shifts from DC to Mobile.

\begin{figure}[t]
    \includegraphics[width=\linewidth]{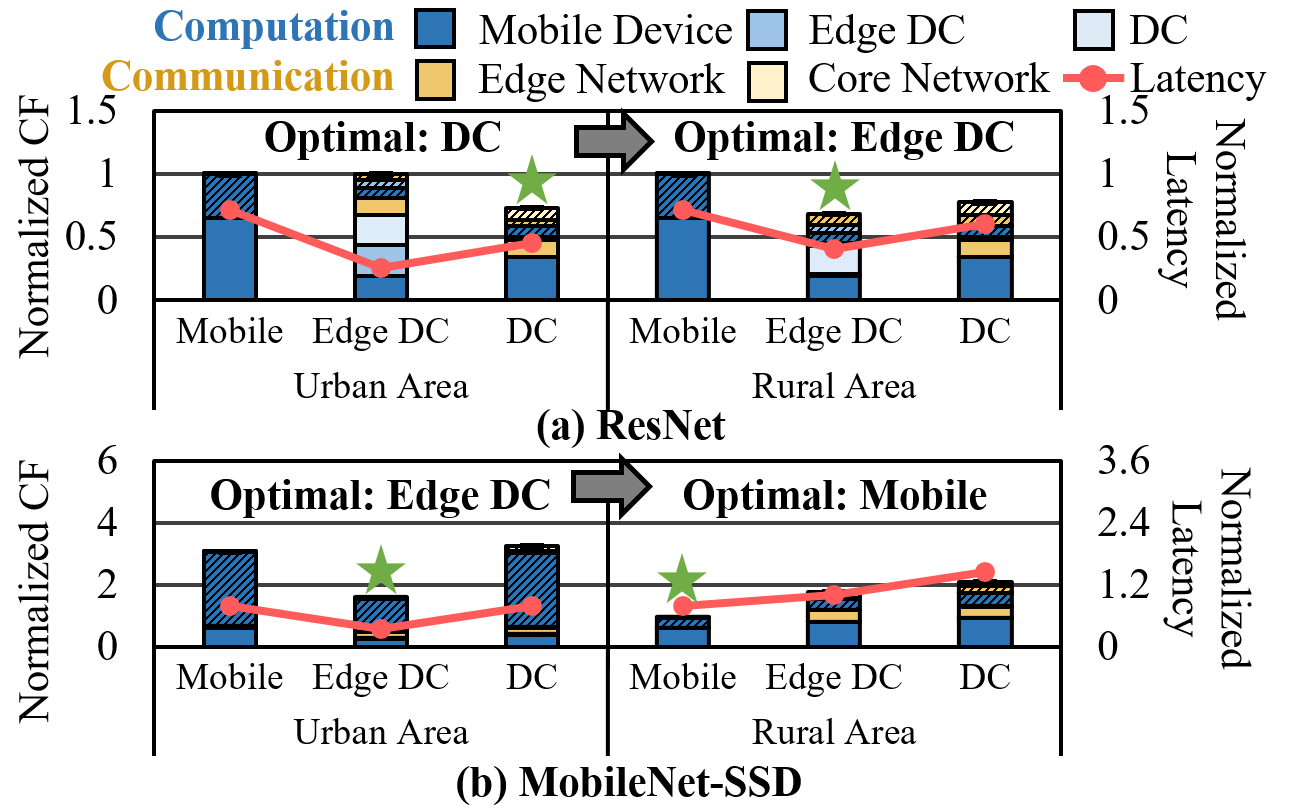} 
    \vspace{-0.25cm}
    \caption{Latency and CF of (a) ResNet and (b) MobileNet-SSD for different carbon intensity scenarios of Edge DC and base station. Note the latency and CF are normalized to those of Mobile in Urban Area. The carbon intensity of Edge DC depends on its location (i.e., Urban Area with less amount of renewable energy and Rural Area with a plenty of renewable energy), but the network latency also depends on the location.}
    \label{fig:geographical}
\end{figure}

\textbf{Carbon intensity of DC: }Fig.~\ref{fig:geographical} shows the latency and CF of (a) ResNet and (b) MobileNet-SSD for different carbon intensity scenarios of Edge DC and base station. The sub y-axis shows the normalized latency. Note the carbon intensity models of mobile device and hyperscale DC are Nighttime Charger and Grid-Mix, respectively. ACT is used as the embodied CF modeling tool, and there is no runtime variance. When the Edge DC and base station are located in an urban area, their carbon intensity will be usually high due to less amount of available renewable energy~\cite{Watttime}. Instead, the edge network latency will be relatively short. On the other hand, when the Edge DC and base station are located in a rural area, the carbon intensity will be low due to a plenty of available renewable energy~\cite{Watttime} --- the edge network latency will increase accordingly. 

Given this geographical trade-off, the carbon optimal execution target depends on the location of the Edge DC and base station, and workload characteristics. In case of ResNet which has a relatively large amount of computations with moderate data transmission overhead, the carbon footprint gain of a Rural Area outweighs the loss from increased network latency (i.e., Edge DC bars). On the other hand, in case of MobileNet-SSD which has a larger transmission data, the latency constraint is not satisfied in the Rural Area due to the increased network latency. This result implies that the selection of the infrastructure components (e.g., Edge DC in Urban Area vs. Edge DC in Rural Area) needs to be based on the careful considerations of geographical trade-off as well as the workload characteristics.

\vspace{0.1cm}
\noindent \textit{An application's carbon cost depends on carbon intensities of electricity used by infrastructures across the entire computing spectrum --- edge, datacenters, and base stations with networking infrastructures connecting the two.}
\vspace{0.1cm}

\begin{figure}[t]
    \includegraphics[width=\linewidth]{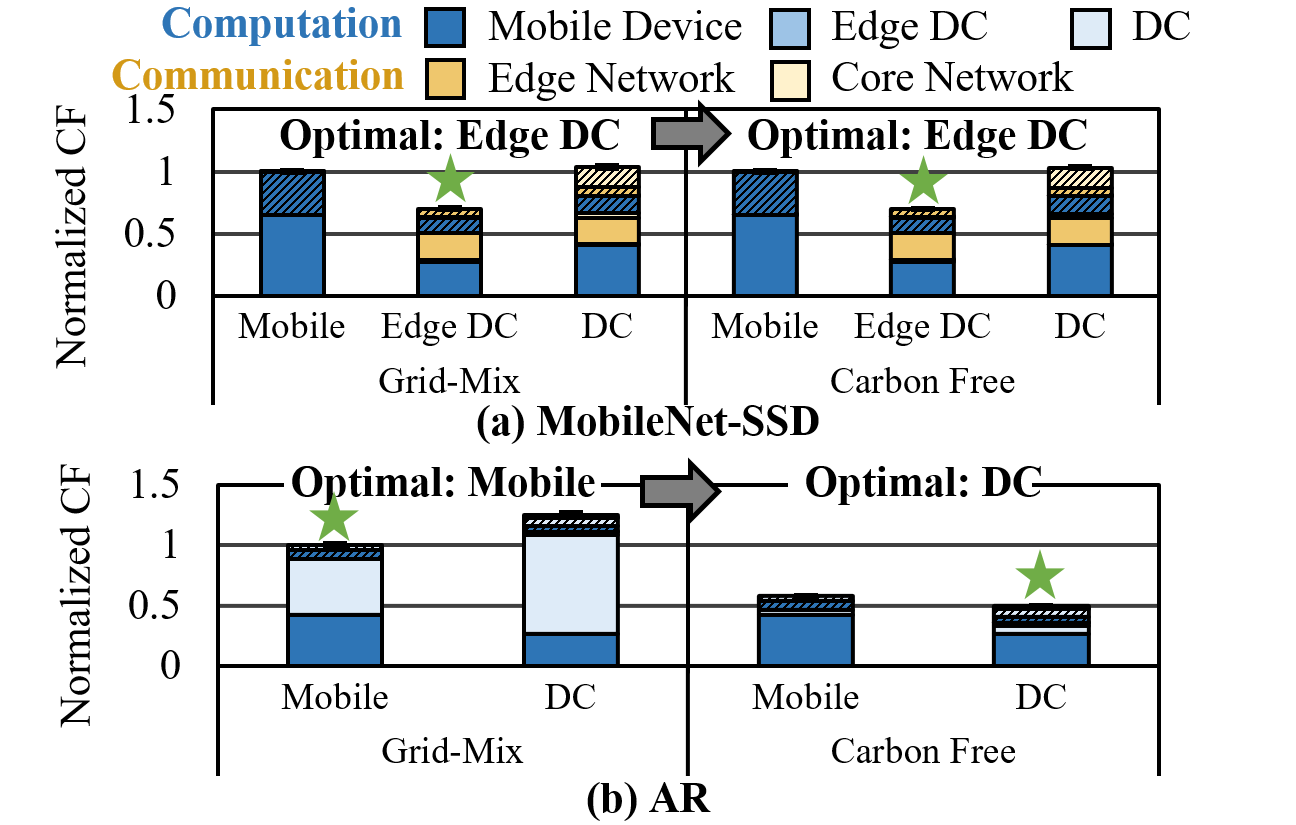} 
    \vspace{-0.25cm}
    \caption{CF of (a) MobileNet-SSD and (b) AR for differnt carbon intensity scnearios of DC. The carbon optimal execution target also shifts with the carbon intensity scenarios (i.e., Grid-Mix and Carbon Free in each column) of DC. The impact of the carbon intensity scenarios of DC on the carbon optimal execution target however varies with workload characteristics.}
    \label{fig:DC CI}
\end{figure}

Fig.~\ref{fig:DC CI} shows the CF of (a) MobileNet-SSD and (b) AR for different carbon intensity scenarios of DC. Note the carbon intensity models of mobile device and edge DC are Nighttime Charger and Urban Area, respectively. ACT is used as the embodied CF modeling tool, and there is no runtime variance. As shown in Fig.~\ref{fig:DC CI}, the impact of carbon intensity on the carbon optimal decision depends on the workloads. In case of MobileNet-SSD, overall CF does not vary significantly even when the energy consumed by DCs is fully covered by renewable energy (i.e., Carbon Free). This is because the operational CF of DC accounts for only a small portion of CF due to the high network overhead and idle overhead of other components. On the other hand, in case of AR, Carbon Free is still beneficial (i.e., DC bars), due to the large amount of computations that can benefit from DC.

\textbf{Uncertainty in carbon intensity: }In a realistic environment, the renewable energy generation can fluctuate even within an hour due to the environmental factors (e.g., clouds turbulence, solar radiation, etc.)~\cite{hirata1995output,kern1989cloud,jewell1987effects,tan2007impact,milan2013turbulent}. Such fluctuations can generate uncertainty in carbon intensity.

To understand the impact of uncertainty in scheduling decisions, we injected the uncertainty into GreenScale based on statistical distributions of renewable energy generation --- according to~\cite{ettoumi2002statistical,celik2004statistical}, fluctuations of solar and wind energy generations can be modeled with beta and weibull distributions, respectively. Even in the presence of 16.8\% of carbon intensity fluctuations, the scheduling decision found by \textit{GreenScale} remain consistent (see error bars in figures). This result implies that the overall conclusions found by \textit{GreenScale} can be generally applicable to realistic environments.

\subsection{Impact of Runtime Variance}
\label{sec:result3}
\vspace{0.4cm}

\begin{figure}[t]
    \includegraphics[width=\linewidth]{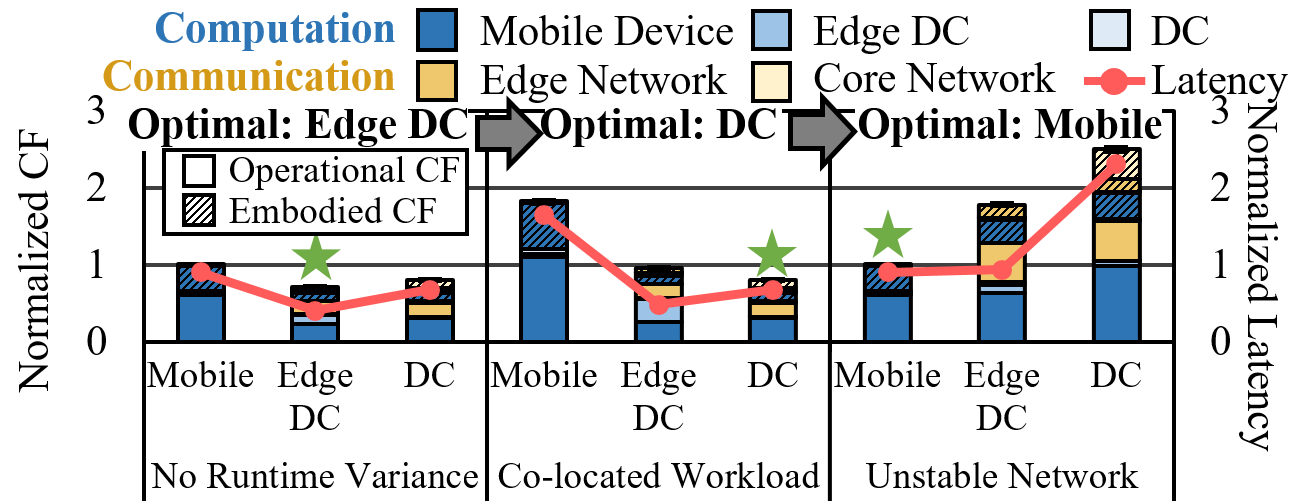} 
    \vspace{-0.25cm}
    \caption{Latency and CF of Inception (a) when there is no runtime variance, (b) when there exist co-located workloads on computer systems, and (c) when the network is unstable. With runtime variance from various sources, the carbon optimal execution target shifts.}
    \label{fig:runtime variance}
\end{figure}

Fig.~\ref{fig:runtime variance} shows the CF and latency of Inception (a) when there is no runtime variance, (b) when there exist co-located workloads, and (c) when the network is unstable. Note the carbon intensity models of mobile device, edge DC, and hyperscale DC are Nighttime Charger, Urban Area, and Grid-Mix, respectively, and ACT is used.

\vspace{0.1cm}
\noindent \textit{Runtime variance, especially common for edge-cloud computing, can impact the carbon cost of an application meaningfully.}
\vspace{0.1cm}

When there is no runtime variance, the carbon optimal execution target is Edge DC. However, when there exist co-located workloads, the carbon-optimal execution target shifts to DC. This is because the adverse impact of co-located workloads depends on the computation and memory capabilities of each component --- DC has the largest computation and memory capabilities among the computing components. On the other hand, when either edge network or core network is unstable, the carbon-optimal execution target shifts to Mobile due to the significantly increased data transmission overhead.


\subsection{Design Considerations of Applications}
\label{sec:result4}
\vspace{0.4cm}

\begin{figure}[t]
    \includegraphics[width=0.96\linewidth]{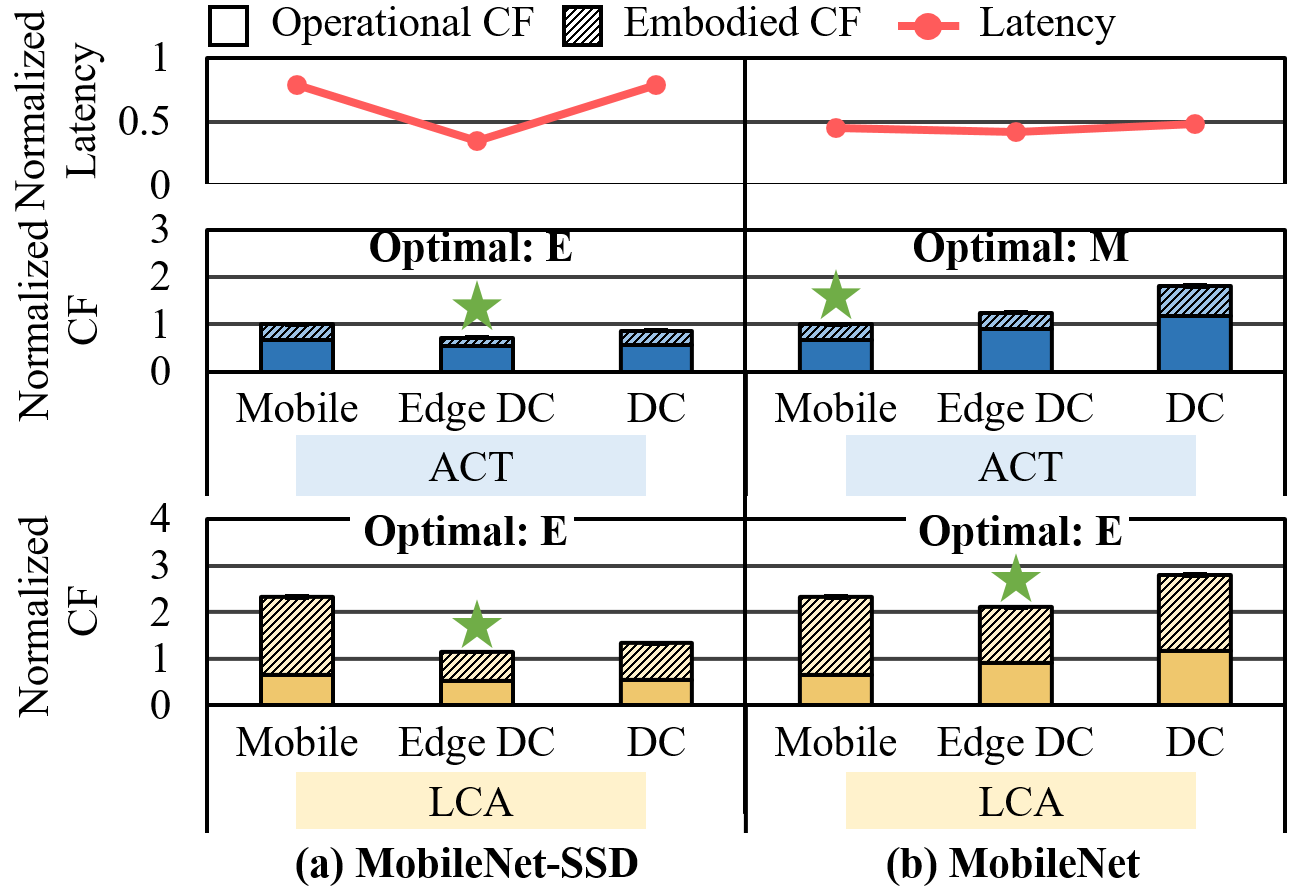} 
    \vspace{-0.25cm}
    \caption{Latency and CF of (a) MobileNet-SSD and (b) MobileNet with two different embodied CF modeling tools. Depending on the amount of estimated embodied CF, the carbon optimal decision can vary.}
    \label{fig:embodied}
\end{figure}

\textbf{Embodied CF model: }Given rising embodied CF of the components, in this study, we explore two embodied CF modeling tools, i.e., ACT and LCA which estimate lower and higher embodied CF of the components respectively. Fig.~\ref{fig:embodied} shows the latency and CF (with two different modeling tools) of execution targets for (a) MobileNet-SSD and (b) MobileNet. Note the carbon intensity models of mobile device, edge DC, and hyperscale DC are Nighttime Charger, Urban Area, and Grid-Mix, respectively, and there is no runtime variance.

In case of MobileNet-SSD, two different modeling tools indicate the same execution target --- Edge DC --- as the carbon optimal execution target. This is because Edge DC shows the shortest latency (which linearly affects the embodied CF) and smallest operational energy at the same time. On the other hand, in case of MobileNet, ACT indicates the Mobile as the carbon optimal execution target whereas LCA indicates the Edge DC as the carbon optimal execution target. This discrepancy comes from different patterns of latency and operational energy consumption of the execution targets --- in case of mobile-friendly MobileNet, operational energy efficiency of Mobile is further optimized. Considering the higher estimation of embodied CF of LCA (compared to ACT), this result implies that \textit{rising embodied CF should be further considered designing green applications in the future}.

\textbf{Number of servers to rent: }Traditionally, the application designers have tried to optimize the number of servers to rent from cloud service providers based on the number of users and their requests. However, in the carbon optimization design space, it is also crucial to consider the efficiency-CF trade-off. To provide a guideline on this parameter, we explore its impact.

\vspace{0.2cm}
\noindent \textit{GreenScale enables application developers to make carbon-informed decisions when provisioning and leasing cloud infrastructures. Carbon-informed cloud resource planning can achieve up to 24.9\% carbon cost reduction.}
\vspace{0.2cm}

\begin{figure}[t]
    \includegraphics[width=0.97\linewidth]{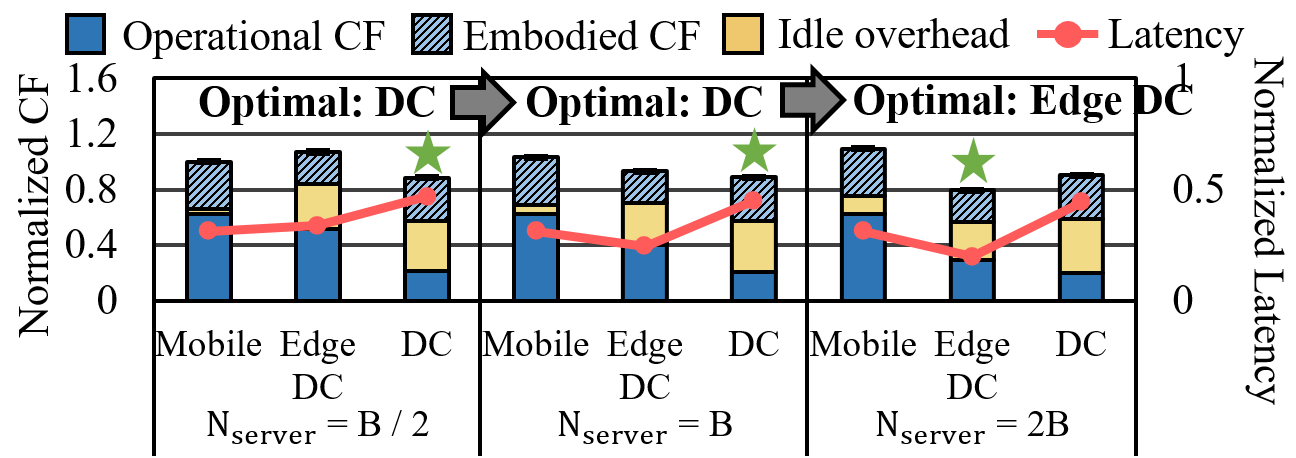} 
    \vspace{-0.25cm}
    \caption{Latency and CF of SqueezeNet when different number of servers are rented from DCs. Latency is normalized to the latency constraint and CF is normalized to Mobile when $N_{servers}$ is $B \slash 2$. Note $N_{servers}$ and $B$ indicate the number of rented servers and the optimal batch size (i.e., 1024), respectively. As the number of rented servers increases (from left to right), the latency and operational efficiency are improved. Due to the improved latency, idle overhead and embodied CF overhead are also improved.}
    \label{fig:servers}
\end{figure}

Fig.~\ref{fig:servers} shows the latency and CF of SqueezeNet when there exist different number of servers in DCs. Note the carbon intensity models of mobile device, edge DC, and hyperscale DC are Nighttime Charger, Urban Area, and Grid-Mix, respectively. ACT is used and there is no runtime variance. As shown in Fig.~\ref{fig:servers}, when the number of servers increases, the latency and operational efficiency are improved. Due to the improved latency, the idle overhead and embodied CF overhead are also improved --- more number of servers are used though. As the improvement is higher in the small-scale edge DC, the carbon optimal execution target shifts from DC to edge DC. 

\textbf{Workload-dependent parameter: }As we present in Section~\ref{sec:result1}, the carbon optimal execution target depends on the computation-communication ratio of workloads. By exploiting workload-dependent parameters, developers can change their computation-communication ratio, improving CF.

\vspace{0.1cm}
\noindent \textit{Application-level optimization levers, such as frame resolution setting (Games) and task partitioning (AR/VR pipeline), provide up to 31.1\% and 14.8\% carbon savings, respectively.}

\begin{figure}[t]
    \includegraphics[width=0.95\linewidth]{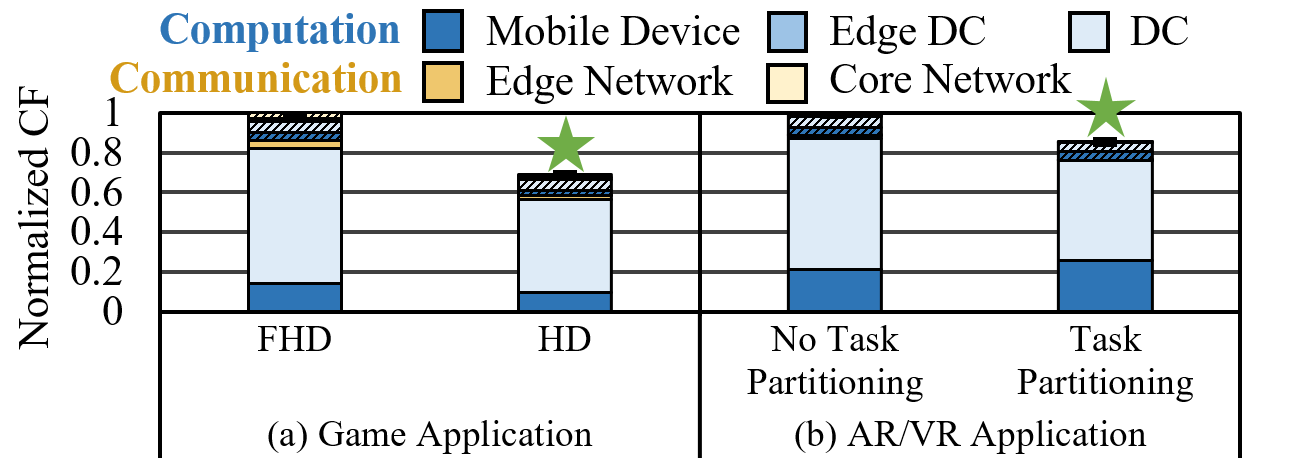}
    \vspace{-0.25cm}
    \caption{By exploiting the workload-dependent parameters, it is possible to change the computation-communication ratio improving the carbon efficiency. Note the carbon intensity models of mobile device, edge DC, and hyperscale DC are Nighttime Charger, Urban Area, and Grid-Mix, respectively. ACT is used as the embodied CF modeling tool.}
    \label{fig:workload}
\end{figure}

In case of the game applications, it is possible to sacrifice the quality of the frames to reduce the amount of computation and data transmission of the execution targets. Fig.~\ref{fig:workload}(a) shows the CF of third-person role-playing game application with different resolution options. When the resolution option is changed from FHD to HD, the overall carbon footprint is reduced (by 31.1\%) along with the decreased computation and data transmission overhead. However, the decreased resolution may have an adverse impact on the QoE of users. 

In case of AR/VR applications, it is possible to partition different pipeline stages between the Mobile and DC by considering the amount of computations required for each stage and the size of intermediate results. Fig.~\ref{fig:workload}(b) shows the CF of AR when we do not consider partitioning the tasks and when we partition the pipeline stages between the Mobile and DC. When tasks are partitioned, the overall carbon footprint is reduced (by 14.8\%). There are two reasons: 1) the size of intermediate results of the stages is lower than that of the input, reducing the data transmission overhead and 2) the utilization of resources (i.e., Mobile Device and DC) increases, reducing the idle carbon footprint by 55.3\%.


\textbf{Scheduling methods: }Since the carbon-optimal execution target varies with various parameters, it is crucial to select a scheduling method that can adapt to the parameters. There have been various scheduling methods that can be used in an edge-cloud execution environment. Since those techniques have different amounts of carbon overhead and accuracy, a careful selection is required. 

\begin{figure}[t]
    \includegraphics[width=0.95\linewidth]{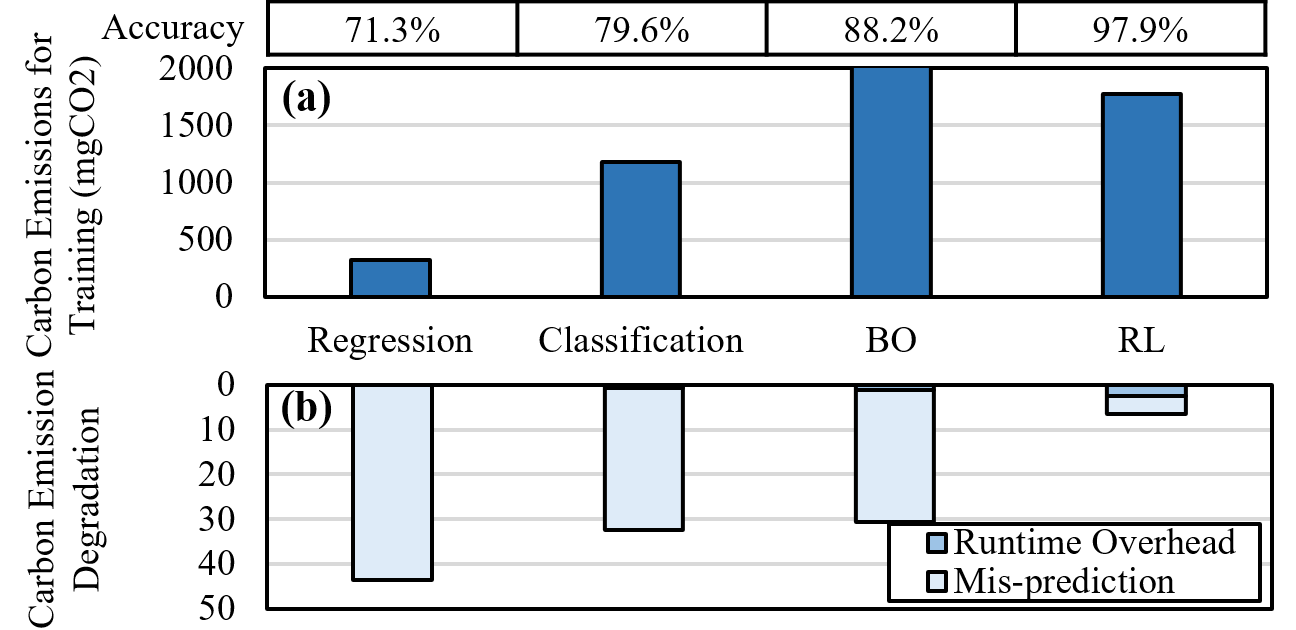} 
    \vspace{-0.25cm}
    \caption{Overhead and accuracy vary across the various scheduling methods (a). Since the runtime overhead and mis-prediction result in the degradation of carbon emissions (b), a careful selection of the scheduling method is required for designing a green application.}
    \label{fig:scheduling}
\end{figure}

To provide the selection guidelines, we custom-design and implement the following scheduling methods, and evaluate their carbon overhead and accuracy\footnote{The accuracy and CF results are obtained using the \textit{GreenScale}, while the overheads are measured using the real-system implementation.} using \textit{GreenScale}: 
\begin{compactitem}
\item Regression which determines the execution targets based on their CF predicted by a regression-based model~\cite{GAFSeber2012}, 
\item Classification which determines the execution targets based on a classification-based model~\cite{JAKSuykens99,BZhang04}, 
\item Bayesian Optimization (BO) which determines the execution targets based on a prediction model obtained by BO~\cite{ASouza2020}, and 
\item Reinforcement Learning (RL) which determines the execution targets using a policy self-learned by considering environmental features~\cite{YGKim2020}. 
\end{compactitem}
Note, for the fair comparison, we fine-tuned all the parameters of the scheduling methods (e.g., hyperparameters of RL).

Fig.~\ref{fig:scheduling} shows (a) the carbon overhead of training and prediction accuracy for the techniques and (b) carbon emission degradation of the techniques compared to the optimal scheduler. In general, the training overhead is one-time cost for the design of the green application, as long as the infrastructure components are not changed. However, prediction accuracy can significantly affect the carbon efficiency of the application since the mis-prediction causes the CF degradation.

In Fig.~\ref{fig:scheduling}(b), the larger runtime overhead is, the smaller carbon emission degradation is, usually coming from the mis-prediction. Among the methods, RL achieves the best accuracy by self-learning a policy to adapt to varying features, such as carbon intensity and runtime variance at the expense of overhead while the others failed to accurately model the non-linear relationship of the features --- the runtime overhead of RL is only 2.4\% of execution time per frame though.

\noindent \textit{GreenScale's carbon-informed decisions can be achieved through a variety of scheduling algorithms with distinct accuracy-overhead trade-offs.}
\vspace{-0.1cm}
\section{Related work}
\vspace{0.5cm}

\textbf{Optimization of edge-cloud applications: }From the perspective of mobile users, optimizing QoE, which is the product of energy efficiency of user-end devices and response time performance, is crucial. Hence, user-centric optimization techniques have tried to identify the best execution target which minimizes the operational energy consumption of the user-end devices satisfying the performance constraint~\cite{ECai2017,AEEshratifar2018,MHan2019,YKang2017,YKim2019,NDLane2016,SWang2020_1,SWang2020_2,GZhong2019}. 

From the perspective of data center operators, optimizing resource utilization, operational efficiency, and execution cost, without affecting the service level agreement (SLA) constraints is crucial. Hence, the cloud-centric optimization techniques have tried to maximize the operational efficiency of infrastructure by splitting the computations across edge-scale and hyperscale servers~\cite{RMo2020, YZhang2018, Ryan2022, STuli2022-2,KKaur2020}, satisfying performance constraints with service level agreements. Among the techniques, several have tried to consider the time-varying renewable energy availability at the DC~\cite{Ryan2022,STuli2022-2}. For instance, \cite{ZLiu2011,ZLiu2011-2} tried to switch compuataions between DCs in different places considering their carbon intensity, whereas \cite{CRen2012} tried to schedule batch jobs considering the renewable energy within a DC. There has been also carbon-aware energy capacity planning work for DCs \cite{RBianchini2012} considering on-site as well as off-site renewable energy generations and their costs. 

Although there have been many techniques, those techniques often fail to make the global carbon optimal decisions, as they do not consider the unique features of carbon optimization, such as varying carbon intensity at the edge, and the amortization of embodied carbon emission. This is mainly due to the absence of framework to quantify and analyze the carbon emissions of the infrastructure components considering the aforementioned features.

\textbf{Carbon modeling tools: }Given the increasing carbon footprint of ICT, academia and industry have proposed a number of tools to quantify carbon emissions across the hardware life cycles. The exergy-based tools follow energy-balance approach to quantify the environmental impact of servers during fabrication and use~\cite{CRHannemann2008,PRanganathan2010}. Although the energy-balance approach can simplify the design space for sustainable systems, the exergy-based tools do not consider the impact of renewable energy during manufacturing and use. The LCA tools quantify the carbon footprint of products across life cycles~\cite{EIO-LCA,SImaPro,Sphera}, not pivoting into the computing systems. LCA tools use coarse-grained information (e.g., economic cost of electronics, system's bill of materials, etc.). Although product environmental reports published by industry have been based on the LCA tools~\cite{Apple2019,Apple-R,TSMC}, they have not been used for comparative analyses between systems or hardware components to guide design space exploration. Complementing the above tools, ACT was proposed to consider the direct carbon footprint from hardware manufacturing and operational use of computing systems~\cite{UGupta2022}. 

Based on the above tools, this work proposes \textit{GreenScale}, a carbon design space exploration and optimization framework considering the unique features of carbon optimization. By using \textit{GreenScale}, we demonstrate carbon optimization is distinct from performance and energy optimization, due to the varying carbon intensity, runtime variance, and embodied carbon emissions. Based on the design space exploration, this work provides a guideline to design \textit{green application}.
\vspace{-0.1cm}
\section{Conclusion}
\vspace{0.5cm}
Given the growing carbon emissions of ICT infrastructure, it is crucial to design \textit{green applications} which can improve the carbon-efficiency of the infrastructure components. In this paper, we propose \textit{GreenScale} a design space exploration and optimization framework. Through the in-depth carbon characterization of state-of-the-art applications across edge-cloud infrastructure, \textit{GreenScale} demonstrates that the carbon optimal scheduling decision depends on various features, such as workload characteristics, \textit{time} and \textit{location}-based carbon intensity, stochastic runtime variance, and the amortization of embodied carbon emissions. Those features make the carbon optimization distinct from performance and energy optimization, leaving significant room to improve carbon-efficiency (by up to 29.1\% compared to a state-of-the-art edge-cloud scheduler).
We believe \textit{GreenScale} can be a viable solution to provide a detailed guideline for developers to design and implement \textit{green applications} that enable sustainable execution.



\bibliographystyle{IEEEtranS}
\bibliography{refs}

\end{document}